\documentclass[twocolumn,showpacs,preprintnumbers,amsmath,amssymb]{revtex4}
\usepackage{epsfig}

\begin{document}


\title{Bending rigidity of stiff polyelectrolyte chains: a single 
chain and a bundle of multichains}

\author{Bae-Yeun Ha}
\email{byha@uwaterloo.ca} 
\affiliation{%
Department of Physics, University of Waterloo, Waterloo, O.N., Canada, N2L 3G1}%
\author{D. Thirumalai} 
\email{thirum@glue.umd.edu}
\affiliation{
Institute for Physical Sciences and Technology, University of Maryland, College Park, MD 20742}%

\date{\today}

\begin{abstract}
We study the bending rigidity of highly charged stiff polyelectrolytes, for both a single chain and many chains forming a bundle.   A theory is developed to account for the interplay between competitive binding of counterions and charge correlations in 
softening the polyelectrolyte (PE) chains.   The presence of even a small concentration of multivalent counterions leads to a dramatic reduction in the bending rigidity of the chains that are nominally stiffened by the repulsion between their backbone charges.   The variation of the bending rigidity as a function of $f_{0}$, the fraction of charged monomers on the chain, does not exhibits simple scaling behavior;  it grows with increasing $f_{0}$ below a critical value of $f_{0}$.  Beyond the critical value, however, the chain becomes softer as $f_{0}$ increases.  The bending rigidity also exhibits intriguing dependence on the concentration of multivalent counterion $n_{2}$; for highly charged PEs, the bending rigidity decreases as $n_2$ increases from zero, while it increases with increasing $n_{2}$ beyond a certain value of $n_{2}$.  When polyelectrolyte chains form a $N$-loop condensate (e.g., a toroidal bundle formed by $N$ turns (winds) of the chain), the inter-loop coupling further softens the condensate, resulting in the bending free energy of the condensate that scales as $N$ for large $N$.  
       
\end{abstract}
\maketitle

\section{Introduction}

Like-charged polyelectrolytes (PEs) such as DNA and F-actin undergo a 
dramatic compaction to highly ordered structures in the presence of 
multivalent 
counterions~\cite{bloomfield,kremer,winkler,pincus,janmey,janmey2,amis,klein,ha.2rods,ha.bundle,ha.pre,ha.angle,Lee,jensen,marcelja,joanny,podgornik,oosawa,manning,oosawa.book,raspaud}.  
Theoretical description of the phases of highly 
charged PE condensates is complicated due to the interplay between the chain deformations and 
counterion-mediated interactions.
This complex interplay is often not treated adequately.  In the 
classical work on stiff PEs~\cite{osf}, the effect of the condensed counterions 
is simply to renormalize the backbone charge density.  More recently,  
Park et al.~\cite{park} and Ubbink and Odijk~\cite{uo} have 
considered the bending rigidity of DNA strands that can form bundles.  The underlying 
assumption in these calculations is that total bending 
energy can be written as a sum over all bending energies from 
individual chains.  It is, however, not obvious when this is valid, since electrostatic 
interactions are long-ranged and the counterion-mediated attractions are not 
pairwise additive.   In a previous study, we 
considered charge fluctuation effects on the bending rigidity of 
randomly charged polymers~\cite{ha.thirum}.  Notably we have shown that charge 
fluctuations along the polymer chains give rise to a reduction in the 
bending rigidity as compared to the corresponding uncharged cases.  
Several other authors~\cite{kardar,andelman,hansen,nguyen} have also 
considered a similar problem of 
the bending rigidity of a single stiff 
PE.  All the existing approaches, however, do not provide much insight 
into a more realistic case, namely bundles of PE chains, since they 
do not take inter-chain coupling into account.  Note that the attraction that 
softens PEs also induces interchain collapse.  
In fact, counterion-induced interactions are not pairwise 
additive and thus the result for a single chain can easily be invalidated 
by the interchain coupling.  
The key to understanding the bending rigidity of PE condensates lies in a consistent 
treatment of both interchain and intrachain couplings.

In this paper, we present a theoretical model for describing  
the bending rigidity of highly charged stiff PEs, both a single 
chain and many chains   
which form bundles.   Throughout this paper, the bending rigidity is 
estimated in terms of the persistence length, which is a 
length scale over which the chain does not change its direction 
appreciably.  The persistence length is simply the bending 
rigidity divided by $k_{B}T$, where $k_{B}$ is the Boltzmann factor and $T$ is the 
temperature.   In particular, we study the 
interplay between preferential adsorption of multivalent counterions 
and charge correlations in determining the bending 
rigidity of PE chains.   We begin by first considering a single chain 
case.   We show that, 
below a certain critical value of the electrostatic strength, the PE
chain becomes stiffer with the increasing 
interaction strength.  Above the critical value, the chain becomes 
softer as the interaction strength increases.  Thus a simple scaling function 
does not describe the bending rigidity.  At first glance, this is quite 
surprising because our results imply that the bending rigidity of 
condensates can be reduced by increasing the backbone charge density.  The 
bending rigidity also 
varies nonmonotonically with the concentration of 
multivalent counterions, $n_{2}$. 
For sufficiently highly charged PEs, the electrostatic bending rigidity drops 
to a large negative value (signalling the collapse of the PE) upon 
a small increase in $n_{2}$ from zero, while it 
decreases in magnitude with increasing $n_{2}$ beyond a certain value of $n_{2}$.  
In other words, these exists an optimal value of $n_2$ at which the PE is most efficiently softened.  In contrast to the previous work~\cite{ha.thirum,kardar}, we consider 
a bundle of multichains as well.  In some cases, a long single PE chain can 
fold back on itself (in a rodlike bundle) or wind around itself (in a 
toroidal bundle) to form a bundle.  In the case of rodlike bundles, 
there is extra complexity arising from hairpin turns of the 
chain at both ends of the bundle, while, in toroidal bundles, 
non-parallel portions of the chain complicate the 
problem~\cite{park}.  Here we 
ignore these comlexities; in the resulting picture, the distinction 
between rodlike bundles and toroidal bundles becomes minor.  With this 
simplification, we consider toroidal bundles without a loss of 
generality.  Whether a bundle is formed by a single 
chain or many different chains, we can consider the bundle to be formed by 
many loops; each loop in a toroidal bundle corresponds 
to one complete turn (wind) of the chain.  As a result, all monomers in a loop 
can be considered as interacting with each other through intra-chain 
(or intra-loop) interactions, since any portions of the loop do not wind around each 
other.  We then discuss the effect of coupling between loops in the 
bundle on the bending rigidity.  Our theory  implies 
that, the bending free energy of each loop in the bundle cannot be 
added up to give the total bending free 
energy.  We also show that the bending free energy cannot be 
computed by summing up the interactions over all pairs of loops.  
The theory rather suggests that the inter-loop coupling further 
softens the condensate.  This is due to the many-body nature of 
electrostatic  interactions induced by counterions.  As a result, the bending free energy of the condensate varies linearly with the number of loops.  
     
\section{Model and Bending Rigidity}

\subsection{general formalism}

Our calculations are based on a model adopted in 
Ref.~\cite{ha.thirum,kardar,andelman}.  The PEs are uniformly 
charged chains, forming a bundle of $N$ loops.   Each loop is assumed to consist of 
$M$ monomers of length $b$, and each monomer carries either a negative charge of $-e$ or 0.  In reality, different loops can have different numbers of monomers, but this polydispersity will not influence the persistence length as long as 
the loop size is somewhat larger than the screening length 
$\kappa^{-1}$, as assumed here.  If $f_{0}$ is the fraction of charged monomers, then each 
monomer is asigned a charge $-e f_{0}$ on average.       
Besides monovalent salts, there are $(Z:1)$ 
multivalent salts as well in the solution.  
Counterions are divided into two 
classes, ``condensed'' ({\i.e.}, they are bound to the PE chain) and 
``free'' ({\i.e.}, they move freely in the solution)~\cite{oosawa,manning}.   The charge on
loop $j$ of monomer $s$  (in units of electronic charge $e$) can assume the 
values
\begin{equation}
q_{j} (s)=-f_{0}^{j} + m_{1}^{j}+ m_{2}^{j} Z
,\end{equation} 
where $m_{1}^{j},m_{2}^{j} =0,1,2,3,...$ are, respectively, the numbers of 
monovalent and multivalent condensed counterions at the site.    Due 
to condensed counterions, the charge distribution along the chain is 
heterogeneous.  The resulting charge destribution is characterized not 
only by the average charge per each site, $\bar q_{j}=\overline{q_{j} 
(s)}=-f_{0}+f_{1}^{j} +Z f_{2}^{j}$, 
but also by the variance in the charge 
per  site $\bar\Delta_{j} = 
\overline{\left[q_{j}(s)\right]^{2}}-\left[{\overline{q_{j}(s)}}\right]^{2} =\left(f_{1}^{j}  + Z^2 
f_2^{j}\right)$~\cite{remark}. Note here that 
$\overline{\left(\ldots\right)}$ is an average 
over all realizations of the charge distribution, and $f_{1}^{j}$ and $f_{2}^{j}$ are 
the average numbers of the monovalent and 
multivalent condensed counterions on loop $j$ per site, respectively.  
The electrostatic interaction Hamiltonian can be expressed in terms 
of $q_{j}(s)$:
\begin{equation}
{{\cal H}_{\rm elec} \over k_{B}T}=
{ \ell_{B} \over 2  } 
\sum_{jj'=1}^{N} \sum_{ss'=1}^{M}  q_{j} (s) 
q_{j'} (s') { {\rm e}^{-\kappa |{\bf r}_{j}(s)-{\bf r}_{j'}(s')|}
 \over |{\bf r}_{j} (s)-{\bf 
r}_{j'}(s')|},
\end{equation}
where  ${\bf r}_{j}(s)$ describes conformations of PE chains 
 in the bundle.  The Bjerrum length $\ell_{B}=e^{2}/4 \pi \varepsilon 
k_{B} T$ is the distance at which the electrostatic interaction between 
two elementary charges $e$ is equal to the thermal energy $k_{B}¥T$, and 
$\varepsilon $ is the dielectric constant of the solution.  The Debye 
screening parameter is  $\kappa^{2} =8 
\pi \ell_{B} I$ where $I$ is the ionic strength.     

The partition function 
is a sum over all charge variables and conformations.  We 
treat the free ions within the three dimensional Debye-H\"uckel 
(DH) 
theory, and consider backbone charges together with their condensed counterions
as forming one-dimensional Debye-H\"uckel (DH) systems~\cite{ha.2rods,ha.bundle}.  
Thus our approach should be contrasted with the approach of Nguyen et 
al.~\cite{nguyen}, which is based on a strongly correlated liquid (SCL) model 
of condensed counterions.
 In Appendix A, we argue that the major contribution to the electrostatic 
 bending rigidity can arise from long-wavelength charge 
 fluctuations as captured by our DH approach, even when the condensed 
 counterions are strongly correlated.  This is because small bending is 
 more efficiently felt by long-wavelength fluctuations, even when 
 counterion ordering is mainly driven by short-wavelength 
 fluctuations.   However, aspects of SCL picture will become more 
 pronounced at low temperatures.  
 
 For a given conformation of a PE described by $[{\bf 
r}_{j}(s)]$, we can 
trace over the charge variables.  If we define the following block matrix 
$\mbox{\boldmath $Q$} [{\bf r}_{j} (s)]$ whose element 
$\mbox{\boldmath $Q$}_{jj'} [{\bf r}_{j} (s)]$ is 
\begin{eqnarray}
\label{matrices}
\mbox{\boldmath$Q$}_{jj'ss'} [{\bf r}_{j} (s)] = {\bar\Delta_{j}}^{-1} 
\delta_{jj'} \delta_{ss'} + 
\ell_{B} {{\rm e}^{-\kappa |{\bf r}_{j} (s)-{\bf r}_{j'} (s')|}  
\over |{\bf r}_{j} (s)-{\bf r}_{j'} (s')|}
,\end{eqnarray}
 then the 
electrostatic free energy functional becomes
\begin{eqnarray}
\label{FC1}
{{\cal F}_{\rm elec} [{\bf r}_{j} (s) ] \over k_{B} T} &=& {1 \over 2} 
\sum_{jj'=1}^{N}
 \sum_{ ss'}^{M}  \bar q_{j}\bar q_{j'}  \Big\{ \bar\Delta_{j}^{-1} \delta_{jj'} \delta_{ss'}\nonumber \\
&& - \bar\Delta_{j}^{-1} \bar\Delta_{j'}^{-1} \mbox{\boldmath$Q$}^{-1}_{jj' ss'} [{\bf r}_{j} (s)]
\Big\}  \nonumber \\
&&+\mbox{$\frac{1}{2}$} 
 \log \Big\{ {\rm 
det}\, \bar\Delta_{j} \mbox{\boldmath$Q$} [{\bf r}_{j}(s)] \Big\}  \nonumber \\
&&+ \mbox{self-energy}
.\end{eqnarray}
Note that the derivation of this result is analogous to the derivation 
of the free energy of bundles of rodlike PE chains which have been studied 
extensively~\cite{ha.2rods,ha.bundle,ha.pre,ha.angle,ha.interface}.   This 
result can readily be obtained by performing standard Gaussian integrals.  So the 
intermediate steps that lead to this result are omitted.  The first term in 
eq~\ref{FC1} corresponds to the screened repulsion,  
while the second term comes from charge fluctuations.  The last term 
comes from $j=j'$ and $s=s'$, and 
is a self energy that should be subtracted from the free energy in 
eq~\ref{FC1}.  Note here that the self energy does not contribute to 
the persistence length, which measures a free energy cost for 
bending or the free energy of the PE chain with reference to a 
rodlike conformation.  This is because the self-energy is the same for all chain 
conformations.         

When PEs are near the rod limit, it is reasonable to assume that they 
are bent uniformly.  The central axis of PEs can then be parameterized by  
${\bf r}_{j} (s)$ as ${\vartheta}_{j}
(s)=s/R$, where $R$ is the radius of curvature.   Note that this 
parametrization amounts to adopting a ground state dominance approximation 
in the chain conformations.    
With this parameterization, the free energy with respect to the reference rod
conformation can be obtained.  To this end, we write
\begin{eqnarray}
\label{matrix.rod}
 \mbox{\boldmath$Q$}_{jj' ss'} &\cong& \mbox{\boldmath$Q$}_{jj' ss'}
[{\vartheta}_{j}(s)]  \nonumber \\
&\equiv&  \bar\Delta_{j}^{-1} \delta_{jj'}  
\delta (R {\vartheta})  + { {\ell_{B}} \over \sqrt{R^{2} {\vartheta}^{2} 
+R_{jj'}^{2} }}\nonumber \\
&&\times {\rm 
e}^{- \kappa \sqrt{R^{2} {\vartheta}^{2}+R_{jj'}^{2} }} 
+ \underline{{\ell_{B}  \over 24} {\rm 
e}^{- \kappa \sqrt{R^{2} {\vartheta}^{2}+R_{jj'}^{2} }} }\nonumber \\
&&\times \underline{\left( { R^{2} {\vartheta}^{4}  \over 
(R^{2} {\vartheta}^{2} +R_{jj'}^{2} )^{3/2} }+ { \kappa 
R^{2} 
{\vartheta}^{4} \over {R^{2} {\vartheta}^{2} +R_{jj'}^{2} } } \right)}
,\end{eqnarray}
where $R_{jj'} $ is the distance between chains $j$ and $j'$.  Note 
that this equation is similar to eq 5.2b in Ref~\cite{ha.thirum}, except 
that the coupling between loops is included in this case.   
It can be shown that the underlined term in 
eq~\ref{matrix.rod} is much smaller than the first two terms near the rod limit, {\i.e.},  
$\kappa R \gg 1$, by the factor of $1 / R^{2}$.  If we denote 
$\mbox{\boldmath$ Q$}_{0} $ to be the corresponding matrix for the 
rod-like case, {\i.e.},  
the first two terms in eq~\ref{matrix.rod}, then we 
can thus write $ \mbox{\boldmath$Q$} 
[{\vartheta}(s)]=\mbox{\boldmath$Q$}_{0} 
\left\{ 1+\left[\mbox{\boldmath$Q$}_{0}^{-1} \Bigl( \mbox{\boldmath$Q$}
[{\vartheta}(s)]-\mbox{\boldmath$Q$}_{0}  \Bigr)\right] \right\}$ and 
consider the term in 
$[\cdot]$ as an expansion parameter.  Convergence of the expansion is 
assured if $\kappa R\gg 1$.  Both 
$\mbox{\boldmath$Q$}^{-1}[{\vartheta}]$ and $\det \mbox{\boldmath$Q$} 
[{\vartheta}]$ can be computed in powers of $1/R^{2}$. 

When PE chains form a bundle of $N$ loops,  charges on 
one loop  correlate {\it not only} with others on the 
same loop {\it but also} with  charges on different 
loops.  In other words, the intra-loop and inter-loop charge fluctuations 
are coupled with each other.  Consistent treatment of 
intra- and inter-loop correlations is hard even when all loops are 
perfectly rigid and parallel (in the absence of conformational 
deformations)~\cite{ha.bundle}.  In the following subsection, 
we will provide explicit solutions for the case of $N=1$.   The 
effects of loop-loop interactions on 
the bending rigidity of PE condensates, namely bundles of PE chains, 
is discussed in Subsection C.    
The electrostatic contribution to the bending free energy of a 
bundle, {\i.e.}, the change in the electrostatic free energy due to 
bending is approximated as 
\begin{eqnarray}
    \label{delF}
{\Delta {\cal F}_{\rm elec} \over k_{B} T} &\cong& {1\over 2} 
\sum_{jj'} \sum_{ss' \in  \vartheta}   {\bar q_{j}\bar q_{j'} \over 
\bar\Delta_{j} \bar\Delta_{j'} }\left[ \mbox{\boldmath$Q$}_{0}^{-1} 
( { \mbox{\boldmath$Q$}}-\mbox{\boldmath$ 
Q$}_{0} )\mbox{\boldmath$Q$}_{0}^{-1}  \right]_{jj'ss'}\nonumber \\
&&
+{1 \over 2} \sum_{j} \sum_{s \in  \vartheta} 
\left[ \left(\mbox{\boldmath$Q$}_{0}^{-1} -\bar\Delta_{j} \right) 
({\mbox{\boldmath$Q$}}-\mbox{\boldmath$Q$}_{0} ) \right]_{jjss } 
,\end{eqnarray}
where the subscripts $j$ and $j'$ run over loops.  
Here, the first term comes from the repulsion between charges, which 
is screened by both condensed counterions and added salt.  The second 
term, which vanishes as $\bar\Delta_{j} \rightarrow 0$, represents the 
attraction due to charge fluctuations in the monomeric charges.   It 
should be noted here that the two terms in eq~\ref{delF} are put on 
an equal footing.  What is suppressed here is non-Gaussian fluctuations 
which are responsible for short-wavelength charge correlations.  We 
argue in the appendix that Gaussian fluctuations can easily dominate 
the $\Delta {\cal F}_{elec}$ as long as $\kappa b$ is small.  This justifies 
our neglect of non-Gausian fluctuations in the computation of $\Delta 
{\cal F}_{elec}$ (thus the persistence length).

When $\kappa^{-1}$ is somewhat smaller than the loop size, then the 
matrix element $\left( \mbox{\boldmath$Q$}_{0}^{-1} \right)_{jj' 
ss'}$ in eq~\ref{delF}, for example,  depends on $|s-s'|$ for given 
$j$ and $j'$.   This enables us to 
further simplify the bending free energy in eq~\ref{delF} by summing it with respect to $s$.  
To this end, let us define matrices $D(s)$ and 
${}^{0}\!D$ as
\begin{eqnarray}
\label{D}
{}^{0}\!D_{jj' } &=& {\xi \over 12 } \int_{0}^{\infty} ds s^{4} {\rm 
e}^{- \kappa \sqrt{s^{2} +R_{jj'}^{2}¥}} \nonumber \\
&&\times \left( {1 \over 
(s^{2}  +R_{jj'}^{2} )^{3/2} }+ { \kappa  \over {s^{2} +R_{jj'}^{2} } } 
\right) \nonumber \\
D_{jj' }(s)
&=&{\xi\over 12} s^4{\rm 
e}^{- \kappa \sqrt{s^{2}¥+R_{jj'}^{2} }}  \nonumber \\
&&\times\left( {1 \over 
(s^{2} +R_{jj'}^{2})^{3/2} }+ { \kappa  \over {s^{2}  +R_{jj'}^{2}} } 
\right),
\end{eqnarray}
and ${}^{0}\!{\cal M}$ and ${\cal M}(k)$ as 
\begin{eqnarray}
\label{M}
{}^{0}\!{\cal M}_{jj'} &=& \bar\Delta_{j}^{-1} \delta_{jj'} +2 \xi K_{0}(\kappa 
R_{jj'} ) \nonumber \\
{\cal M}_{jj'} (k) &=& \bar\Delta_{j}^{-1} \delta_{jj'} +2 \xi 
K_{0}(R_{jj'} \sqrt{\kappa^{2} +k^{2} })
,\end{eqnarray}
where $\xi \equiv \ell_B/b$ 
and $K_{0}(x)$ is the zeroth-order modified Bessel function of the second kind.  
In terms of ${}^{0}\!D,D(s),{}^{0}\!{\cal M}$, and ${\cal M}(k)$, the 
bending free energy can be partly diagonalized.   In other words, the 
free energy can be Fourier transformed from $s$ to its Fourier 
conjugate $k$. 
To order $1/R^{2}$, we obtain   
\begin{equation}
{ \Delta{\cal F}_{\rm elec} \over k_{B} T} \approx 
{L \over 2 R^{2} } \ell_{elec}  
,\end{equation}
where
\begin{eqnarray}
\label{lelec}
\ell_{elec} &\simeq& {1 \over b} 
 \sum_{jj'} {\bar q_{j} \bar q_{j'} \over  \bar\Delta_{j} \bar\Delta_{j'} }  
 \left[ \left({}^{0}\!{\cal M}\right)^{-1} \, {}^{0}\!D
\, \left({}^{0}\!{\cal M}\right)^{-1} \right]_{jj'}  \nonumber \\
&&+ \sum_{jj'=1}^{N} \int_{0}^{\infty} ds D_{jj'} (s) \int {dk \over 2 \pi} 
 [{\cal M}^{-1}(k) 
-\bar\Delta ]_{jj'} \nonumber \\
&& \times\cos ks 
,\end{eqnarray}
where $\bar\Delta_{jj'} =\bar\Delta_{j} \delta_{jj'}$.  
Despite its apparent complexity, the expression in eq~\ref{lelec} can 
readily be derived in the same spirit as the bundle free 
energy~\cite{ha.bundle,ha.pre,ha.interface}.  In Appendix B, we outline the detailed intermediate mathematical steps that lead to eq~\ref{lelec}.  Note that 
the bending free energy cannot be written as a sum of two-body  
interactions over all pairs of loops.  This is consistent with the  
earlier finding that rod-rod interactions in polyelectrolyte 
solutions are not pairwise additive~\cite{ha.bundle,ha.pre}.  It is not surprising to 
see that the breakdown of pairwise additivity is also manifested in the bending 
free energy.  

\subsection{single-loop cases: $N=1$}

 For $N=1$, the bending free energy 
 given in eq~\ref{lelec} can be further 
 simplified:
\begin{eqnarray}
\label{elec.bend.torus.single}
\ell_{elec} & \simeq& 
{\ell_{\rm OSF} \over \left[1+2 \xi \bar\Delta  K_{0}(\kappa b) 
\right]^{2} } - 
{(\xi \bar\Delta)^{2} \over 12}\int ds  
(s^{2}+s^{3} \kappa){{\rm e}^{-\kappa s} \over s} \nonumber \\
&&\times \int {dk \over 2 \pi} {2  K_{0} (b \sqrt{k^{2}+\kappa^{2}})\over 
1 + 2 \xi\bar\Delta   K_{0}(b \sqrt{k^{2}+\kappa^{2}})}\cos ks 
.\end{eqnarray}
In this equation and in what follows we drop the subscript 1 in 
$\Delta_{1}$ if it refers to a single-loop case.  
This expression should be compared with the OSF result, {\i.e.},  
$\ell_{\rm OSF} 
=\ell_{B} \bar q^{2} /4 b^{2}\kappa^{2}$, where $e\bar q$ is the ``renormalized'' 
charge per site and, in our notation, is $\bar q=(-f_{0}+f_{1}+Zf_{2})$.   
In OSF theory, the 
effect of condensed counterions 
is simply to reduce uniformly the backbone charge density.  In contrast to OSF result, 
the effects of condensation 
cannot be described in terms of a single renormalized 
parameter $\bar q$.  The first term in 
eq~\ref{elec.bend.torus.single}, arising from the net charge 
repulsion, tends to stiffen the chain, while  the second term, 
originating from  
charge fluctuations, softens the chain and is responsible for the chain 
collapse at low temperatures~\cite{kremer,pincus,klein}.  

In the limit of  small $\kappa b$,  
the persistence length is 
dominated by large-scale charge fluctuations, corresponding to 
small-$k$ contributions.  This leads to the following asymptotic 
expression for  the persistence length 
\begin{equation}
    \label{ell.single.asymp}
    \ell_{elec} \simeq {\ell_{\rm OSF} \over (1+2 \xi \bar\Delta \, 
K_{0}(\kappa b) )^{2} } - {\xi^{2}\bar\Delta^{2}   \over 16} 
{\kappa^{-1} \over 1+ 2 \xi \bar\Delta  K_{0} (\kappa b) }
.\end{equation}
Note that the charge correlation contribution, {\i.e.,} the second 
term,  does not exhibit simple 
scaling behavior as a function of $x\equiv \bar\Delta \xi$.  For small 
$x$, it varies quadratically with $x$, but it crosses over 
to one that scales as $x$ for large $x$.

\begin{figure}[ht]
\centering
\epsfig{file=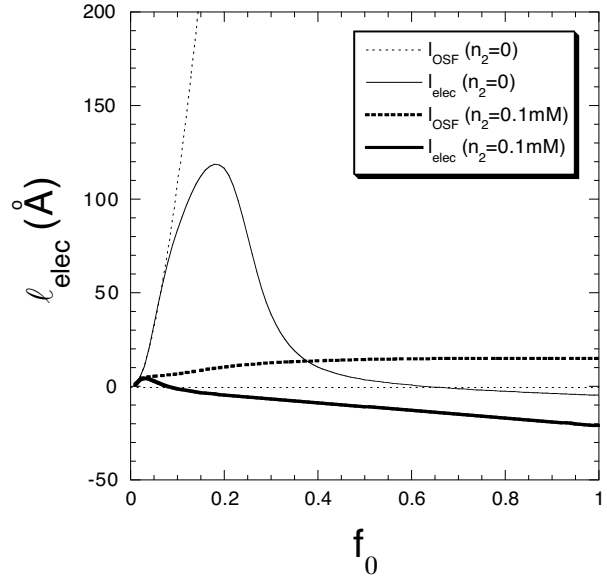,width=3.2in}
 \caption{The electrostatic persistence length $\ell_{elec}$ (the  
solid lines) as a function of $f_{0}$ (the fraction of charged monomers).  We have chosen the parameters  
$b=1.7{\rm \AA}$, $\ell_{B}=7.1{\rm \AA}$, and $n_{1}=1{\rm mM}$.   
Due to the competition between the net charge repulsion and charge correlations, 
$\ell_{elec}$ does not vary monotonically.    
In the absence of multivalent ions $(n_{2}=0)$, $\ell_{elec}$ grows 
monotonically with $f_{0}$ up to 
$f_{0}=0.2$, consistent with OSF result (the thin dotted line).  
Beyond this, however, charge correlations become important and tend to 
shrink the persistence length.    In the presence of a small 
concentration of 0.1mM of trivalent 
counterions ($Z=3$),  charge correlation effects are dominant and 
tend to reduce $\ell_{elec}$ beyond $f_{0} \simeq 0.05$.  
Over a wide range of $f_{0}$ ($f_{0} > 0.1$), $\ell_{elec}$ is negative.   Due 
to preferential adsorption of multivalent counterions, even OSF 
result (the thick dotted line) deviates significantly from the 
corresponding monovalent case (the 
thin-dotted line)  }
\label{lelec.vs.f0}
\end{figure}

The OSF result is valid for $\bar\Delta
\xi  \ll 1$ (Cf. Fig.~\ref{lelec.vs.f0}).  
Thus when the charge fluctuations are 
important, OSF result is not accurate and the fluctuation 
correction should be incorporated properly, as will also be detailed later.  
Our results in 
eq~\ref{elec.bend.torus.single} should be compared with those for 
polyampholyte chains 
reported before~\cite{ha.thirum}.  In the 
former case, we sum up all multipole expansions~\cite{remark''}, while in the 
latter case the free energy is expanded in powers of $\bar q$ and 
$\bar \Delta$.  Inclusion of all multipole 
terms~\cite{ha.bundle,ha.pre} is 
crucial in the present case.  
This is because the counterion-mediated attraction becomes important 
when the chain is highly charged, {\i.e.}, for large values of 
$f_{0}$.  In this case, the multipole expansion 
diverges~\cite{ha.bundle,ha.pre}.  
On the other hand, $q$ and $\bar\Delta$ can be adjusted {\it independently} in the case of polyampholytes, and there can be finite ranges of parameters where the 
perturbative description is valid.  The convergence of the 
expansion is ensured for the screened cases if $\bar\Delta$ is small 
enough, as is the case for the problem discussed in Ref~\cite{ha.thirum}.  
More recently, Golestanian et al.~\cite{kardar} and Ariel and Andelman~\cite{andelman}  have also considered a similar problem of the bending rigidity of a stiff polyelectrolyte.   In contrast to the theory presented 
here, they only consider single-loop cases.  At a low salt limit, our 
attractive bending rigidity  scales as  $ \kappa^{-1} \log (1/\kappa b)$, which 
is in agreement with their result.  On the other hand, our result in 
eq~\ref{ell.single.asymp} differs from the one in Ref.~\cite{nguyen}, which is 
based on a SCL model of condensed counterions.  If our result includes 
charge correlations driven by long-wavelength fluctuations, then the 
SCL approach captures short-wavelength charge correlations.  Note that 
these two kinds of charge correlations should contribute to 
$\ell_{elec}$.  More precisely the competition between the two will 
essentially determine the leading behavior of the persistence 
length.   In Appendix A, we provide a heuristic argument to study 
this competition.  We find that the persistence length can be mainly 
determined by long-wavelength correlations as captured by our approach, 
even when the charge correlation between counterions is dominated by short-wavelength correlations.  This is because small bending is more efficiently 
felt by long-wavelength charge correlations.   In other words, the mode of charge 
correlations that dominates $\ell_{elec}$ also depends on the mode of chain 
deformation, not just on counterion ordering.

The electrostatic free energy in eq~\ref{FC1} depends on the 
average number of the condensed counterions per monomer, 
$f_{\alpha}$ $(\alpha=1,2)$.  To solve for
$f_{\alpha}$ self-consistently, we equate the chemical
potentials of condensed and free ions~\cite{oosawa.book}.  In the following derivation, 
the subscript $\alpha=1$ and 2 refer to monovalent and multivalent ions, 
respectively.    The chemical potential of free 
ions is mainly associated with the configurational entropy of mixing and is approximately given 
by $\mu_{\alpha}^{free} /k_{B} T \simeq \log n_{\alpha} v_{0}$, where 
$n_{\alpha}$ is the 
bulk ion concentration and $v_{0}$ is the volume of ions.  
The chemical potential of condensed counterions arises from the 
attraction of counterions to PE chains and the entropic penalty for 
confinement~\cite{remark3}: 
\begin{equation}
    \label{mu.cond}
    {\mu_{\alpha}^{cond} \over k_{B}T} \simeq -  Z_{\alpha} \xi \bar q K_{0} (b \kappa) + 
\log f_{\alpha}+ {\mu_{\alpha}^{fluc}\over k_{B}T}, \quad \alpha=1,2
,\end{equation} 
where $\mu_{\alpha}^{fluc}$ is the charge fluctuation 
contribution to the chemical potential given by 
\begin{eqnarray}
   && {\mu_{\alpha}^{fluc} \over k_{B}T} = \left({\partial \over \partial 
f_{\alpha} }\right) \bigg\{ b \int_{0}^{\infty}{dk \over 2 \pi} \Big[ 
\log \Big( 1 + 2 \bar \Delta \xi \nonumber \\
&&\, \times K_0 (b \sqrt{\kappa^2+k^2})\Big)- 
2 \bar\Delta \xi K_0 (b \sqrt{\kappa^2+k^2})  \Big] \bigg\}
.\end{eqnarray} 
Note that the term in $\{\ldots\}$ is a standard 
charge fluctuation free energy of an one-dimensional Debye-H\"uckel 
system immersed in a three dimensional ionic fluid~\cite{ha.bundle}.   Strictly speaking, our expression for  $\mu_{\alpha}^{cond}$ is 
valid for a rodlike chain conformation.  In principle, the effect of chain 
deformation can be included; $\mu_{\alpha}^{cond}$ can also be expanded 
in powers of $1/R$.  The leading-order behavior of $\ell_{elec}$, however,  is not 
influenced by the chain deformation correction to $\mu_{\alpha}^{cond}$, since it 
only leads to a subleading correction to $\ell_{elec}$.     The equilibrium 
value of $f_{\alpha}$ can be fixed by 
requiring $\mu_{\alpha}^{free} =\mu_{\alpha}^{cond}$.  Note that $\mu_{\alpha}^{fluc}$ 
can be important when $Z$ is large and $\kappa^{-1}$ is 
finite~\cite{nguyen1}.

To demonstrate the potency of multivalent counterions in softening PE 
chains, we have solved for $f_{\alpha}$ and the electrostatic persistence length 
$\ell_{elec}$ simultaneously.  
We plot $\ell_{elec}$
 as a function of $f_{0}$ in Fig.~\ref{lelec.vs.f0}.  
 We have chosen the parameters 
$n_{1}=1{\rm mM}$, $b=1.7 {\rm \AA}$, $r_{c}=2 {\rm \AA}$, 
and $\ell_{B}=7.1{\rm  \AA}$ (corresponding to $T=300$K in water for which 
$\varepsilon =80$).   First consider the case for which 
counterions are monovalent ({\i.e.}, $n_2=0$) as described by the thin curves.  
In this case, $\ell_{elec}$ changes non-monotonically as 
$f_{0}$ increases.  Our calculation should be compared with the 
corresponding OSF 
result $\ell_{OSF}$, which varies monotonically.  Note that the OSF curve 
does not vary quadratically with $f_{0}$ beyond $f_{0} \simeq 0.2$.  This 
is because condensed counterions start to renormalize the backbone charge 
density beyond $f_{0} \simeq 0.2$.  The discrepancy between OSF 
result and ours may seem surprising.  Our theory predicts that the chain size 
increases as the strength of Coulomb 
interaction increases, up to $f_{0} \simeq 0.2$.  Beyond this, the chain 
shrinks its size with the increasing backbone charge fraction.  
This puzzling behavior can be understood in terms of the competition 
between the net charge repulsion and counterion-mediated 
attractions.  When $f_{0} < 0.2 $, the long-ranged repulsion is dominant 
and the size of chain grows approximately quadratically with $f_{0}$, 
consistent with OSF result.  The OSF behavior crosses over to the 
attractive regime where the charge fluctuations start to shrink the 
chain size.   For sufficiently large $f_{0}$, $\ell_{elec}$ can 
become negative.  In other words, the bending rigidity of highly 
charged PE chains is smaller than the corresponding non-ionic chains.  
The reduction in the bending rigidity is attributed to the charge 
correlation effect that becomes dominant over the respulsive 
contribution due to a finite excess charge.  

The effect of counterion condensation is far more pronounced in the presence of 
a small concentration of 0.1mM of trivalent counterions ($Z=3$).  
Both OSF result (the bold dotted line) and our result (the bold solid 
line) start to deviate from the monovalent case (corresponding to the thin 
curves) beyond $f_{0} \simeq 0.05$.  In other words, the presence of 0.1mM 
of trivalent counterions is more influential on the bending rigidity 
of PE chains than that of 1mM of monovalent ions.  The result for 
$\ell_{elec}$ in this case is strikingly different from the 
corresponding monovalent case; $\ell_{elec}$  is 
much smaller than in the corresponding monovalent  case as long as 
$f_{0} > 0.05$; it is also negative over a wider range of $f_{0}$.   
Our results clearly demonstrate the efficiency of 
multivalent counterions in shorterning the persistence length of the PE 
chain, thus causing the collapse of the PE chain.  The efficiency of 
$Z=3$ in reducing $\ell_{elec}$ can be attributed to the interplay between 
preferential adsorption of multivalent counterions (over monovalent 
ones) and charge fluctuation effects in determining the persistence 
length; highly charged macrions can preferentially bind multivalent 
counterions even when the bulk concentration is dominated by 
monovalent ions.  Thus multivalent counterions are far more efficient 
on a molar basis in neutralizing the macroion charge~\cite{jacob}.  
The OSF 
result in the presence of a tiny concentration of multivalent 
counterions can strongly deviate from that for monovalent cases.  
Additionally, the strength of charge fluctuations grows approximately 
linearly with the valency of counterions, if preferential 
adsorption is assumed.  When these two 
effects are combined, multivalent counterions can dramatically change the 
bending rigidity of PE chains as demonstrated in Fig.~\ref{lelec.vs.f0}.  

\begin{figure}
\centering
\epsfig{file=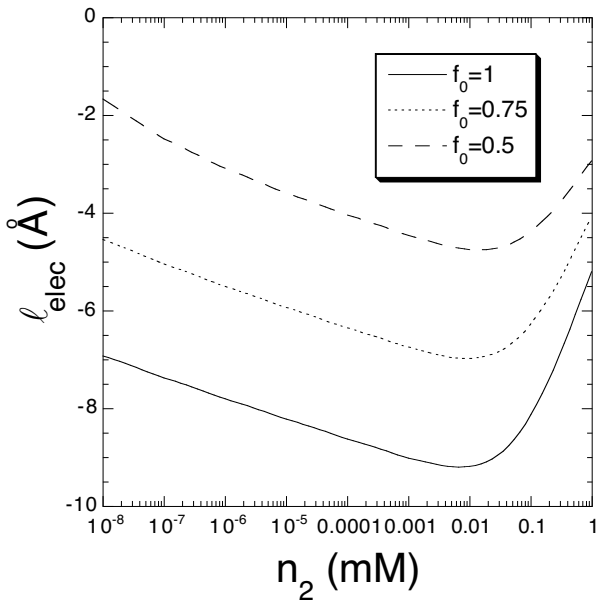, width=3.2in}
 \caption{The electrostatic persistence length (described by the 
solid line) as  a function of $n_{2}$ (the concentration of $Z$-valent counterions) for $b=1.7\AA$, $\ell_{B}=7.1\AA$, $n_{1}=1 {\rm mM}$, 
and $Z=3$ (trivalent salts).  The persistence length $\ell_{elec}$ varies 
non-monotonically with $n_{2}$ for the all cases of $f_0=1,0.75,0.5$ and takes on a minimum at a certain value of $n_2$ in the range $0.01 {\rm mM} <n_2 < 0.1 {\rm mM}$; only a small concentration is needed to soften PEs.   
The efficiency of trivalent counterions in softening the PE chain arises from the interplay between preferential adsorption of 
multivalent counterions and charge fluctuations in determining $\ell_{elec}$.  
Also note that there is an optimal 
value of $n_{2}$ at which $\ell_{elec}$ has the minimum value.  
At zero concentration of PE chains as in this figure, only a tiny concentration of 
multivalent counterions 
is needed to almost completely neutralize the backbone charges of the PE chain.  
Beyond the optimal value, multivalent ions start to contribute 
toward screening the charge correlation effect, leading to the 
non-monotonic variation of $\ell_{elec}$ with $n_{2}$.  
\label{lelec.vs.n2}}
\end{figure}

To further demonstrate the dramatic effects of multivalent counterions, we 
have estimated the electrostatic persistence length as a function of the 
concentration of multivalent counterions $n_{2}$.  We 
have chosen the parameters $Z=3$, $n_{1}=1{\rm mM}$, $b=1.7 {\rm \AA}$, $r_{c}=2 {\rm \AA}$, and $\ell_{B}=7.1 {\rm \AA}$.  We plot the electrostatic persistence 
length $\ell_{elec} $
in Fig.~\ref{lelec.vs.n2} as a function of $n_{2}$ for a few different values of $f_0$.  Our results in the figure are 
intriguing.  For all these cases ({\i.e.}, $f_0=1,0.75,0.5$),  $\ell_{elec}$ 
varies non-monotonically with $n_{2}$.  It becomes more negative  as $n_2$ increases from zero up to a certain value of $n_2$.  Beyond this, it decreases in magnitude with increasing $n_2$.   
This implies that there is an optimal concentration at which $\ell_{elec}$ is most negative.  Somehow this optimal value is not sensitive to $f_0$; it falls in the range $0.01 {\rm mM} \le n_2 \le 0.1{\rm mM}$ and hence only a small concentration is needed to soften PE chains.  This clearly demonstrates the efficiency of multivalent counterions in softening oppositely charged PE chains.  For the entire range of $n_2$ adopted in Fig.~\ref{lelec.vs.n2}, $\ell_{elec}$ does not show a simple scaling law.  When $n_2$ is somewhat smaller than the optimal value, $\ell_{elec}$ has weak dependence on $n_2$ as shown in the figure.  We find that, for all cases ({\i.e.}, $f_0=1,0.75,0.5$), $\ell_{elec}$ assumes the simple scaling form:  $\ell_{elec} \sim c + d  \log n_2$, where $c$ and $d$ are $n_2$-independent negative numbers and the concentration $n_2$ is in mM.  
As evidenced in Fig.~\ref{lelec.vs.n2}, $d$ ({\i.e.}, the slope of the curves) is insensitive to $f_0$;  it has been estimated to be close to $-0.2{\rm \AA}$ for all three cases.  In contrast, $c$ changes linearly with $f_0$ and is more negative for larger $f_0$; for the chosen parameters, $c$ has shown to be approximately given by $c \approx -10 f_0$ (\AA).  It should be emphasized that both $c$ and $d$ are non-universal constants that can depend on $\kappa$ and $Z$ for fixed $\ell_B$ and $b$.   Also note that this simple scaling behavior can be invalidated by the competitive binding on the one hand and the presence of the opposing effects in determining $\ell_{elec}$ (cf. eq~\ref{elec.bend.torus.single}) on the other.      


The efficiency of multivalent counterions 
as evidenced by our results in Fig.~\ref{lelec.vs.n2} can also be 
understood in terms of competitive binding and charge fluctuations.  
When $n_{2}=0$, only the monovalent 
counterions bind to the PE chain.  When $n_{2} \ne 0$, however, 
the PE chain can preferentially bind multivalent counterions, 
replacing monovalent counterions in its vicinity, thus softening the PE 
chain, as can be seen by solving the linear equations in 
eq~\ref{mu.cond}
simultaneously~\cite{remark1}.  
For sufficiently highly charged PE chains $(f_{0} \simeq 1)$ at zero 
concentration, only a tiny  
concentration of multivalent counterions is needed to replace 
monovalent counterions.  
 Note that the optimal value of $n_{2}$ at 
which $\ell_{elec}$ has the minimum value depends not only on $n_{1}$ 
but also on the concentration of PE chains.  Beyond the optimal value, 
multivalent counterions start to contribute toward screening, weakening 
the correlation effect and thus leading to the non-monotonic variation  
of $\ell_{elec}$ with $n_{2}$.  
Our result in Fig.~\ref{lelec.vs.n2} corresponds to PE chains at zero 
concentration and thus a direct comparison of our result to experimantal 
data for PE chains at non-zero concentration should be made with due caution.  
Finally, computer simulations have explicitly provided 
evidence for the efficiency of multivalent counterions in inducing 
collapse of strongly charged polyelectrolytes~\cite{Lee}.

 \subsection{Multi-loops}

So far we have restricted ourselves to the single-loop case.  Due to 
the breakdown of the pairwise additivity of charge-fluctuation 
interactions~\cite{ha.bundle,ha.pre}, it is important to discuss the effects of 
inter-loop coupling on the bending rigidity.  Based on our common intuition, 
we expect the inter-loop interactions to stiffen the 
condensate.  This is because the net charge repulsion is longer 
ranged than the attraction~\cite{ha.2rods}.  
 This conjecture is, however, based on the 
pairwise additivity of the electrostatic interactions between 
different loops and can easily be invalidated unless the charge 
fluctuations are small enough~\cite{ha.bundle,ha.pre,ha.interface}.  
For the case of parallel rods, it has been 
shown that the pairwise 
additivity is valid only when the interaction is 
repulsive~\cite{ha.bundle,ha.pre}.   By the same 
token, it is crucial to include non-pairwise additive interactions or 
many-body effects~\cite{thirum} in the description 
of the bending rigidity of PE condensates.  Because the solution of 
eq~\ref{lelec} in general cases is difficult, we invoke physically 
motivated  simplification.  To understand the physical consequences of 
non-pairwise additive interactions on the 
bending rigidity, we make some approximations.  First we do not 
explicitely include excluded volume repulsion between loops.  To keep 
different loops from approaching arbitrarily close to each other, we assume that loops in 
the condensate are arranged on a square lattice with lattice 
constant $a$.  The lattice constant $a$ can be considered as an 
equilibrium distance between two neighboring loops, which is 
determined by the balance of a few competing effects: charge 
correlation attractions, excluded volume repulsions, etc.  We also assume 
that all loops in the bundle remain parallel with each other, even when they 
bend, and that $\bar\Delta_{j}$ and $\bar q_{j}$ are independent of $j$ and taken to be equal to $\bar{\Delta}$ and $\bar q$, respectively.  With this simplification, we can take advantage 
of this periodicity to recast the problem using discrete Fourier transform.  The 
resulting  bending free energy is given by
\begin{eqnarray}
    \label{lelec.fourier}
\ell_{elec} &\simeq& { \bar q^{2} \, \left(  {}^{0}\! \tilde{D} \right)
 \over  \left[1+2 \xi \bar{\Delta} \, \left({}^{0}\tilde{K_{0}}\right)\right]^{2}}   - \sum_{j_{1} 
 j_{2}=0}^{\sqrt{N}-1} 
\int ds  D_{{\bf j}_{\bot}} (s)  \nonumber \\
&& \times \sum_{{\bf 
 k}_{\bot}} \int {dk \over 2 \pi}  {2 \bar{\Delta} \xi \tilde{K_{0}}({\bf k}_{\bot},k)\over 
 1+ 2 \bar{\Delta} \xi 
\tilde{K_{0}}({\bf k}_{\bot},k) } \cos ({\bf k}_{\bot} \cdot {\bf j}_{\bot})\nonumber\\ 
&& \times \cos ks
,\end{eqnarray}
where ${\bf k}_{\bot}$ is the wave vector conjugate to $a {\bf 
j}_{\bot }$, where ${\bf j}_{\bot} \equiv (j_{1}, j_{2})$ and $j_{1}, 
j_{2} =0.1,\ldots, \sqrt{N}-1$,  and is given by ${\bf 
k}_{\bot} = {2 \pi \over \sqrt{N}} (n_{1},n_{2})$ with $n_{1}, 
n_{2}=0,1, \ldots , \sqrt{N}-1$.  We denote the discrete Fourier 
transform of a function $g (a | {\bf j}_{\bot} | 
\sqrt{\kappa^{2} +k^{2} })$ as
\begin{equation}
\tilde g ( k,{\bf k}_{\bot} ) \equiv 
\sum_{j_{1} j_{2}=0}^{\sqrt{N}-1} g (a | {\bf j}_{\bot} | 
\sqrt{\kappa^{2} +k^{2} }) \cos {\bf k}_{\bot} \cdot {\bf j}_{\bot}
.\end{equation}
Finally, the function ${}^{0}\tilde g$ is simply $g (k=0,{\bf 
k}_{\bot} =0)$.

First note that the bending free energy of $N$-loop condensates  is 
not simply $N$ times that of a single loop.  Our theory 
suggests that each loop in a bundle is further softened due to the 
inter-loop coupling, as implied by eq~\ref{lelec.fourier}.  We expect that  the 
repulsive bending rigidity per loop $decreases$ with $N$.  This is because 
coupling between charges on different loops enhances screening in 
electrostatic repulsions.  Similarly, the attractive interaction
becomes stronger as $N$ increases, due to enhanced charge 
correlations as will be detailed later.  More precisely, inter-loop correlations enhance 
intra-loop correlations.  A similar issue for the case of rigid PEs is 
discussed by Ha and Liu~\cite{ha.correlation}.  
The implication of eq~\ref{lelec.fourier} 
is most {\it striking} in the limit of $\kappa b \rightarrow 0$:  
For sufficiently large $N$, the repulsive term in 
eq~\ref{lelec.fourier} scales as $1/N$ and becomes 
vanishingly small as $N \rightarrow \infty$.  Thus, only the attraction 
can modify the bending rigidity in this case.      
 Recently, 
Ha and Liu~\cite{ha.interface} have considered two interacting bundles 
of randomly-charged rods and 
shown that, as $\kappa b\rightarrow 0$, the interfacial interaction comes 
from 
charge-fluctuation attractions only. This is analogous to the vanishingly 
small contribution of the repulsion to the bending rigidity in the 
present  
case.  For the screened case of $\kappa \ne 0$, on the other hand, the 
repulsion also contributes to the bending rigidity. 

For sufficiently large $N$, we can show that the condensate has a 
well-defined bulk bending free energy: $\Delta {\cal F}_{\rm elec} 
\sim V \ell_{bulk}$ where $V$ is the condensate volume and $\ell_{bulk} $ is 
$N$-independent persistence length.  To establish this, we examine the 
behavior of ${}^{0} K_{0}, \tilde{K_{0}} (k,{\bf k}_{\bot} )$, and 
$\tilde{D} (k,{\bf k}_{\bot})$ for large $N$.  For large $x$, 
$K_{0}(x) \sim 1/\sqrt{x}  {\rm e}^{-x}$ and thus 
${}^{0}\tilde{K_{0}},{}^{0}\tilde{D}, \tilde{K_{0}} (k,{\bf k}_{\bot} )$ approach  
constants as $N \rightarrow \infty$.  Similarly, it can be shown 
that ${}^{0} \tilde{D},  \tilde{D} (k,{\bf k}_{\bot})$ also 
become constants as $N \rightarrow \infty$.  This is trivially true 
for finite $\kappa$.  As $\kappa \rightarrow 0$, it suffices to 
establish this asymptotic behavior for 
the attractive term only, since the repulsive term becomes dominated in 
this limit.  As $\kappa \rightarrow 0$, $D_{ij}(s)$ in eq~\ref{D} 
vanishes 
{\it at least} as fast as $1/R_{ij}^{2}$ for large $|i-j|$.  Thus 
the descrete Fourier 
transform is convergent.  Furthermore, for large $N$, we can 
replace the sum $\sum_{{\bf k}_{\bot} }$ can be replaced by 
$N \int_{0}^{2 \pi/a} {d{\bf k}_{\bot} \over (2 \pi)^{2}} $, 
up to a correction of order unity.  From these 
arguments, we conclude that the bending free enegy per monomer 
increases linearly with $N$ for sufficiently large $N$.   This 
enables us to write the total bending free energy as 
$ \Delta {F}_{elec}/k_{B}T= {1 \over 2} \ell_{bulk}  V/R^{2}a^{2}$, where 
$\ell_{bulk} $ is $N$-independent persistence length per loop. 
An asymptotic behavor of $\ell_{bulk}$ can be obtained replacing the 
summations by integrals; we found that, for $\kappa D \ll 
1$, $\ell_{bulk}$ varies as 
\begin{equation}
\ell_{bulk}\sim -{\kappa^{-1}  \over \log (\kappa^{-1}/ D)}
,\end{equation}
where $D$ is the crosssectional diameter of the bundle, $D=\sqrt{N} 
a$.  This should be compared with the corresponding result for a 
single loop case; the PE chains are further softened by inter-loop 
couplings by the factor of $\log (b/D)$.  
Park et al.~\cite{park} have used a similar expression for the bending free 
energy 
(Cf. their eq 1) without the benefit of derivation.      
Here, we have 
shown that ``many-body effects'', {\i.e.}, loop-loop couplings,
 lead to the bending free energy of PE condensates which grows 
 linearly with $N$ for $N\gg1$, providing a quantitative basis for the previous 
 work of Park et al.~\cite{park}.    
 
\section{Conclusions}

In conclusion, we have studied the effect of charge fluctuations 
on the stiffness of highly charged stiff PE chains.  In particular, we have 
focused on the interplay between competitive binding of counterions
to PE chains and charge correlations between condensed 
counterions in softening the PE chains.  
We have shown that the bending rigidity of the PE chain, in the 
presence of counterion condensation,   
cannot be captured by a simple scaling behavior.  The existence of multiple distinct  
regimes characterizes the conformations of highly charged 
PEs.  This is a consequence 
of the simultaneous presence of 
competing interactions, which tends to invalidate a simple 
scaling analysis. 
Our theory also illustrates the significance of non-pairwise 
additivity of counterion-mediated interactions in softening PE chains; we have shown 
that the inter-loop coupling enhances softening, resulting in a 
well-defined bulk bending free energy.    
   
Acknowledgements: This work was supported in part by the National 
Science Foundations through grant number CHE0209340 (DT) and by the 
Natural Science and Engineering Research Council of Canada (BYH).  We are grateful to D. Andelmann for helpful comments.

\section*{Appendix A}

In this appendix we argure that the Debye-H\"uckel (DH) approach may 
be valid over a much wider parameter space than implied by  
a simple thermodynamic consideration in Ref.~\cite{nguyen}.  In 
Ref.~\cite{nguyen}, condensed counterions are considered as forming a 
strongly correlated liquid (SCL) confined to the surface of their 
binding PE chains.  It was then argued that the negative, charge 
correlation contribution to the persistence length is dominated by these 
strong charged correlations of 
the SCL.  In this appendix, we argue the thermodynamic argument 
underestimates the importance of long-wavelength charge fluctuations 
as compared to SCL correlations, since it ignores the coupling of charge 
correlations to bending.  Recently, it has been shown that, at high 
temperatures, long-wavelength (LW) charge fluctuations dominate the free 
energy, while short-wavelength (SW) fluctuations are dominant at low 
temperatures~\cite{ha.mode}.  Even below the freezing temperature, there exists a 
LW contribution to the free energy.  Interestingly the major contribution to 
the electrostatic bending rigidity of a polyelectrolyte chain can arise from 
LW charge fluctuations even when the 
free energy is dominated by SW fluctuations, as long 
as the chain is near the rod limit.  This is because small bending is 
more effectively felt by the LW charge fluctuations.  In this case the 
DH approach ought to be a good approximation.  

To focus on the essential physics of this issue, we here consider a 
single-loop case  where all backbone charges are neutralized 
by  counterions.  
Near the rod limit, we can write the 
interaction Hamiltonian as follows: ${\cal H}={\cal H}_{\rm rod} + 
\Delta {\cal H}$ where ${\cal H}_{\rm rod}$ corresponds to the rodlike 
conformation and $\Delta {\cal H}$ is the change in the interaction 
due to bending.  The  free energy cost due 
to bending is
$$
 \Delta {\cal F} = \left< \Delta {\cal H} \right>=
 {1 \over 2} \sum_{ss'} D(s,s') \left< q(s) q(s') \right> +{\cal O}\left( R^{-4}\right)
 ,\eqno(A1)$$
where $\left<\cdots\right>$ is an average with respect to the Boltzmann 
factor ${\rm e}^{-{\cal H}_{\rm rod}/k_{B} T} $ and $D(s,s')$ is 
$$
D(s,s') ={\xi \over 24}   {{\rm e}^{- 
\kappa |s-s'|} \over |s-s'|} \left( |s-s'|^{2}+ \kappa  |s-s'|^{3} \right) 
.\eqno(A2) 
$$
The charge correlation contribution to the persistence length can be read 
 off from this expression:
 $$
 \ell_{corr} = {\xi \over 24 L} \sum_{ss'}D(s,s')
\left< q(s) q(s')\right>
.\eqno(A3)$$
 Now the computation of 
the persistence length reduces to the computation of the charge 
correlation function $\left< q(s) q(s')\right>$.  So far our formalism 
is exact up to $1/R^{2}$.  
 
 If we use a Debye-H\"uckel (DH) approximation, then $\left< q(s) q(s') 
 \right>$ is simply 
 $$
 \left< q(s) q(s') \right>_{DH}  = Q^{-1} (s,s')
 ,\eqno(A4)$$
 where $Q$ is a matrix whose matrix element $Q (s,s')$ is 
 $$
 Q(s,s') = \bar\Delta^{-1} \delta_{ss'}  + \ell_{B} {{\rm e}^{-\kappa 
 |s-s'|}\over |s-s'|}
 .\eqno(A5)$$
As a result, the persistence length in 
 eq~$A3$ reduces to the one in eq~\ref{ell.single.asymp}.  In the 
 limit of $\kappa \rightarrow 0$, 
 the DH approximation leads to 
$$
\ell_{DH} \simeq - {\left(\xi \bar\Delta \right)^{2} \over 16}{\kappa^{-1}\over 
1+2 \xi \bar\Delta \log(1/\kappa b)} 
.\eqno(A6)$$
The persistence length in this expression grows linearly with the Debye 
screening length.
 
 At low temperatures, the backbone charges together with the condensed 
 counterions tend to be strongly correlated.  To simplify the problem 
 we assume that $f_{0}=1$ and that counterions are divalent ($Z=2$) and form an ionic crystal such 
 that the charge distribution along the chain is represented by a ground 
 state $(+-+-+-+-)$.  The resulting charge 
 correlation can then be approximated by the following oscillatory 
 function: $\left< q(s) q(s') \right>_{osc} \simeq 
 \cos (\pi |s-s'|/ b)$.  If we use this expression in 
 eq~$A3$, then we have
 $$
 \ell_{osc}  = {\xi\over 6 } \sum_{s=1}^{\infty}  {  {\rm e}^{- 
\kappa s} \over s} \left( s^{2}+ \kappa s^{3} \right) \cos \left( {\pi 
s \over b} \right) 
. \eqno(A7)
$$
In the limit of $\kappa b \rightarrow 0$, $\ell_{osc}$ approaches 
$$
    \label{ell.osc}
\ell_{osc} =-{\ell_{B} \over 26} 
.\eqno(A8)$$
This result can be compared to the one based on the SCL model of 
polyelectolytes~\cite{nguyen}, $\ell_{SCL} = - Z^{2} \ell_{B} /96$.   For 
$Z=2$, this leads to $\ell_{SCL} = -\ell_{B}/24$.   Note 
that $\ell_{SCL}$ is only slightly different from the one in 
eq~$A8$.  This 
difference can be attributed to the fact that we treat the charge fluctuations 
differently from the SCL approach.  Note that $\ell_{DH}$ is much larger 
than $ \ell_{SCL}$ for small $\kappa b$.  This is becase the LW charge 
fluctuations  are much more effectively felt by small bending in this 
case.  

However, neither of $\ell_{DH}$ nor $\ell_{OSC}$ solely describes the bending rigidity of 
polyelectrolyte chains for a wide range of $T$ or $f_{0}$.  This is because $\ell_{OSC}$ 
suppresses fluctuations  while $\ell_{DH}$ does not accurately capture 
strong charge fluctuations.  In general both LW and SW charge 
flcutuations contribute to the persistence length.  To study the 
competition between the two, we use a linear response theory and 
assume that the free energy of the charge fluctuations on a rod can be 
written in the Fourier space as
$$  
{{\cal H}_{\rm rod} \over k_{B} T}={1 \over 2} \sum_{k} \delta q (k) S^{-1} 
(k) \delta q (-k)
,\eqno(A9)$$
where $S(k)$ is the charge structure factor.  The probability of the 
fluctuation $\delta q(k)$ is proportional to the factor 
$\exp\left[- S^{-1} (k) \right]$.  At high temperatures, $S(k) \equiv 
S_{DH} (k) \simeq Q^{-1} (k)$ while 
at low temperatures $S(k) \equiv S_{OSC} (k) \simeq \delta_{k \, {\pi 
\over b}} $.  At intermediate temperatures, we assume that $S(k) \sim A(T) S_{DH} (k) +B (T) S_{OSC}(k)$.   Within this approximation, the charge 
correlation persistence length is given by 
$$\begin{aligned}
\label{deltaell}
 \ell_{corr} \sim& - A(T)  \ell_{DH} - B(T)  \ell_{OSC} \nonumber \\
\sim&  -A(T) {\xi \bar\Delta \over 32  }{\kappa^{-1} \over \log (1/\kappa b)} - 
B(T){\ell_{B} \over 26}
\end{aligned}
.\eqno(A10)
$$
Suppose the free energy is dominated by short-wavelength 
fluctuations, {\i.e.}, $A (T) \ll B (T)$.  Even in this case, however, 
the persistence length can be dominated by long-wavelength 
fluctuations, {\i.e.}, $\ell_{corr} \sim A(T) \ell_{DH} $ in 
the limit of $\kappa b \rightarrow 0$.    
This is because small bending is much more effectively felt by 
the long-wavelength fluctuations in this limit.  Unfortunately, the precise form of $A(T)$ and $B(T)$ for a wide range of $T$ are not known to date.  Nevertheless it is clear that $A (T) \approx 1$ is a good approximation at high temperatures, while $B(T) \rightarrow 1$ as $T \rightarrow 0$.  Near and at the crossover region between $\ell_{DH}$ and $\ell_{OSC}$, however, $A(T)$ and $B(T)$ may deviate from this asymptotic value, {\i.e.}, 1.  If we use this, we would get a simple criterion for the DH approach to be valid:
$$
{\kappa^{-1}      \over \log \left( 1/\kappa b\right) } > 1
.\eqno(A11)
$$
Note that this can easily be satisfied at a low salt limit.  This implies that the electrostatic persistence length can be mainly determined by $\ell_{DH}$, even when $A(T) \approx B(T)$.   As a result, our DH approach is valid for a much wider parameter space than implied by Ref.~\cite{nguyen} which suppressed the interplay between chain deformation and charge correlations in determining the mode of charge correlations that dominates the persistence length.

\section{Appendix B}

In this appendix, we outline the derivation of our 
major result in eq~\ref{lelec}.  In this appendix we use $\mbox{\boldmath$D$}$ to  denote $\mbox{\boldmath$Q$}-\mbox{\boldmath$Q_0$}$.   First consider the first term of 
eq~\ref{lelec}.  To simplify this term, consider   
$$
\begin{aligned}
  I &\equiv  \sum_{ss'}\left[ \mbox{\boldmath$Q$}_{0}^{-1} 
\mbox{\boldmath$D$} \mbox{\boldmath$Q$}_{0}^{-1}  \right]_{jj'ss'}  \nonumber \\
&= \sum_{ss's''s'''} \left[ {\left( \mbox{\boldmath$Q$}_{0}^{-1}\right)}_{ss''} 
 \mbox{\boldmath$D$}_{s'' s'''} {\left(\mbox{\boldmath$Q$}_{0}^{-1} 
\right)}_{s''' s'} \right]_{jj'}
\end{aligned} 
.\eqno(B1)
$$
In this equation and in what follows, $\left( \mbox{\boldmath$A$}_{ss'}\right)_{jj'} = \mbox{\boldmath$A$}_{ss'jj'}$, where $\mbox{\boldmath$A$}$ is a block matrix such as $\mbox{\boldmath$Q$}$.  It then follows that $\left[ \left(\mbox{\boldmath$A$} \mbox{\boldmath$B$}\right)_{ss'}\right]_{jj'}=\left( \sum_{s''} \mbox{\boldmath$A$}_{ss''} \mbox{\boldmath$B$}_{s''s'}\right)_{jj'}=\sum_{j''}\sum_{s''} \mbox{\boldmath$A$}_{jj''ss''} \mbox{\boldmath$B$}_{j''j's''s'}$, where $\mbox{\boldmath$B$} $ is also a block matrix of the same rank.    
Note that $\left( \mbox{\boldmath$Q$}_{0}^{-1}  \right)_{jj''ss''}$ 
is a function of $|s-s''|$. 
It is easy to show that  
$$
    \sum_{s} \left( \mbox{\boldmath$Q$}_{0}^{-1}\right)_{jj'' ss''} = 
\left[ \sum_{s}\left(\mbox{\boldmath$Q$}_{0}^{-1}\right)_{ss''}\right]_{jj''} ={}^0\!{\cal M}^{-1}_{jj''}
,\eqno(B2)
$$
where ${}^0\!{\cal M}$ is the same matrix defined in eq 8.  Here, the second step follows from 
$$
\begin{aligned}
\left[ \sum_{s} \left( \mbox{\boldmath$Q$}_{0}^{-1} \right)_{ss''} \right]_{jj''}&= \left[ \sum_s \left( {1\over \bar\Delta^{-1} + \mbox{\boldmath$V$} }\right)_{ss''}\right]_{jj''} \nonumber \\
&=\left[ \sum_s \bar\Delta \left( 1- \bar\Delta \mbox{\boldmath$V$} + \bar\Delta^2 \mbox{\boldmath$V$}^2 + ...\right)_{ss''}\right]_{jj''}\nonumber \\
&= {1\over b}\left[ \bar\Delta \left( 1- \bar\Delta 2 \xi K_0 + \Delta^2 (2 \xi K_0)^2+... \right)\right]_{jj''}\nonumber \\
&= {}^0\!{\cal M}^{-1}_{jj''}
\end{aligned}
,\eqno(B3)
$$
 where $\mbox{\boldmath$V$}_{jj'ss'}=\ell_B {e^{-\kappa |{\bf r}_j(s)-{\bf r}_{j'}(s')|}\over |{\bf r}_j(s)-{\bf r}_{j'}(s')|}$ and $\left(K_0\right)_{jj'}=K_0 (\kappa R_{jj'})$.   
If we use this in eq $B1$, then we have
$$
  I = \left[ {}^0\!{\cal M}^{-1}
\sum_{s'' s''' }\mbox{\boldmath$D$}_{s'' s'''} {}^0\!{\cal M}^{-1} \right]_{jj'}
.\eqno(B4)$$  
This , if combined with eq 6, essentially leads to the first term of eq 10 in the continuum limit.    

To simplify the second term of eq 10, consider
$$
J \equiv \left( \mbox{\boldmath$Q$}_{0}^{-1}  
\mbox{\boldmath$D$} \right)_{jjss }
=\left[ \sum_{s'} \left(\mbox{\boldmath$Q$}_{0}^{-1} \right)_{ss'}\mbox{\boldmath$D$}_{s's} \right]_{jj}
.\eqno(B5)
$$
Here $\left(\mbox{\boldmath$Q$}_{0}^{-1} \right)_{ss'}$ can be expanded as follows:
$$
\begin{aligned}
\left(\mbox{\boldmath$Q$}_{0}^{-1} \right)_{ss'}&=\left( { 1\over \bar\Delta_j^{-1} + \mbox{\boldmath$V$}  }  \right)_{ss'}\nonumber\\
&= \bar\Delta_j \bigg( \delta_{ss'} -\bar\Delta_j \mbox{\boldmath$V$}_{ss'}+ \bar\Delta_j^2 \sum_{s''} \mbox{\boldmath$V$}_{ss''}\mbox{\boldmath$V$}_{s''s'}+...\bigg)
\end{aligned}
.\eqno(B6)$$
Using this in eq $B5$, we have
$$
\begin{aligned}
J =&\bar\Delta_j \bigg[  \mbox{\boldmath$D$}_{ss} -\sum_{s'} \bigg( \bar\Delta_j \mbox{\boldmath$V$}_{ss'}\mbox{\boldmath$D$}_{ss'} \nonumber \\
&-\bar\Delta_j^2 \sum_{s's''} \mbox{\boldmath$V$}_{ss''} \mbox{\boldmath$V$}_{s''s'} \mbox{\boldmath$D$}_{s's} \nonumber \\
&+ \bar\Delta_j^3 \sum_{s's'' s'''} \mbox{\boldmath$V$}_{ss''}\mbox{\boldmath$V$}_{s'' s'''} \mbox{\boldmath$V$}_{s''' s'}\mbox{\boldmath$D$}_{s's}+...\bigg)\bigg]_{jj}
\end{aligned}
.\eqno(B7)$$
Note that the first-order term in $\bar\Delta_j$ in this equation will cancel the term containing $\bar\Delta$ in the second term of eq 10.  As a result, the second term of eq 10 becomes
$$\begin{aligned}
\mbox{2nd term}&=-{1 \over 2} \sum_j  \bar\Delta_j \bigg[  \sum_{ss'} \bigg( \bar\Delta_j \mbox{\boldmath$V$}_{ss'}\mbox{\boldmath$D$}_{ss'} \nonumber \\
&-\bar\Delta_j^2 \sum_{s''} \mbox{\boldmath$V$}_{ss''} \mbox{\boldmath$V$}_{s''s'} \mbox{\boldmath$D$}_{s's} \nonumber \\
&+ \bar\Delta_j^3 \sum_{s'' s'''} \mbox{\boldmath$V$}_{ss''}\mbox{\boldmath$V$}_{s'' s'''} \mbox{\boldmath$V$}_{s''' s'}\mbox{\boldmath$D$}_{s's} +...\bigg)\bigg]_{jj}
\end{aligned}
.\eqno(B8)$$

In the Fourier space $k$, conjugate to $s-s'$, this equation can be diagonalized with respect to $s$ and $s'$ and resummed: 
$$\begin{aligned}
\mbox{2nd term} =& -{M \over 2} \sum_j  \bar\Delta_j \bigg[  \int{dk \over 2\pi} \bigg( \bar\Delta_j \mbox{\boldmath$V$}(k)\mbox{\boldmath$D$}(k) \nonumber \\
&-\bar\Delta_j^2  \mbox{\boldmath$V$}(k) \mbox{\boldmath$V$}(k) \mbox{\boldmath$D$}(k) \nonumber \\
&+ \bar\Delta_j^3 \mbox{\boldmath$V$}(k)\mbox{\boldmath$V$}(k)\mbox{\boldmath$V$}(k)\mbox{\boldmath$D$}(k) +...\bigg)\bigg]_{jj'} \nonumber \\
&= -{M \over 2} \sum_j  \bar\Delta_j \bigg[  \int{dk \over 2\pi} {\bar\Delta_j \mbox{\boldmath$V$}(k) \mbox{\boldmath$D$}(k)\over 1+ \bar\Delta_j \mbox{\boldmath$V$}(k) } \bigg]_{jj}
\end{aligned}
,\eqno(B9)$$
where $f(k)$ is a Fourier transform of $f(s)$: $f(k)=\sum_s f(s) \cos ks$.
This essentially leads to the second term of eq 10.





\begin{references}
\bibitem{remark}  Note that $\bar\Delta_{j}$ is related to the
concentration (or number) fluctuations of condensed counterions in chemical 
equilibrium with free counterions.  First of all, we should 
emphasize that the average involved in the estimation of this 
qualtity is different from the esemble average denoted by $\left< 
\ldots\right>$, which is an average 
over all realizations of $q_{j}(s)$, weighted by the Boltzmann 
factor: 
\begin{equation}
    \left<\ldots\right>= {\overline{\left[ \ldots \exp\left(-{\cal 
H}_{elec}/k_{B}T\right)\right]} \over  \overline{\left[\exp\left(-{\cal 
H}_{elec}/k_{B}T\right)\right]}}
.\end{equation} 
In other words,  $\left< \ldots \right>$ reduces to $\overline{\left(\ldots\right)}$ if 
the electrostatic interactions among counterions are turned 
off.  In this case, we can consider condensed counterions as 
forming an ideal gas in chemical contact with free counterions.  
As a result, charges at different sites are uncorrelated. 
If $M^{j}_{1}$ and $M^{j}_{2}$ 
are the total number on loop $j$ of monovalent and multivalent 
counterions, respectively, then it follows from a simple thermodynamic 
consideration that 
\begin{equation}
    \overline{\left( M_{\alpha}^{j}\right)} -\left( 
    \overline{M_{\alpha}^{j}}\right)^{2} =\overline{M_{\alpha}^{j}¥}
    ,\end{equation}
where $\alpha=1,2$.   This gives rise to charge fluctuations along the 
chain; the total-charge fluctuation of condensed 
counterions is 
\begin{equation}
M \Delta_{j}=\sum_{\alpha=1}^{2} Z_{\alpha}^{2}\left[ ¥ 
\overline{\left( M_{\alpha}^{j}\right)} -\left( 
    \overline{M_{\alpha}^{j}}\right)^{2} \right] = 
    \sum_{\alpha}¥Z_{\alpha}^{2} \overline{M_{\alpha}^{j}¥}
    ,\end{equation} 
    where $Z_{1}=1$ and $Z_{2}=Z$.  This leads to the variance in the 
    charge per site charge used in Sec. II: $\Delta_{j} = f_{1}+Z^{2} f_{2}$.
\bibitem{remark''} Note here that the multipole expansion refers to 
an expansion of the free energy (thus eq~\ref{ell.single.asymp}) in powers 
of $\bar\Delta$ 
within the one-dimensional Debye-H\"uckel (DH) approach.  Truncation of high order 
terms in this expansion can be justified only when $f_{0}$ (thus 
$\bar\Delta$) is small enough.  In this sense we have 
summed up the free energy over all realizations of charge 
fluctuations (within the DH approach).     As a result, non-Gaussian 
fluctuations are suppressed.  The validity of this approach 
is discussed in the appendix, which implies that the DH approach is 
valid even for large values of $f_{0}$ as long as $\kappa b$ is 
sufficiently small.
\end{references}
\end{document}